\documentclass[fleqn,usenatbib]{mnras}

\usepackage{newtxtext,newtxmath}
\usepackage[T1]{fontenc}
\usepackage{ae,aecompl}

\usepackage{graphicx}   % Including figure files
\usepackage{amsmath}    % Advanced maths commands
\usepackage{amssymb}    % Extra maths symbols
\usepackage{color}
\usepackage{caption}
\usepackage{subcaption}
\title[Spatial intensity interferometry...]{Spatial intensity interferometry on three bright stars}
\author[W. Guerin et al.]{
W. Guerin$^{1}$\thanks{E-mail: william.guerin@inphyni.cnrs.fr},
J.-P. Rivet$^{2}$\thanks{E-mail: rivet@oca.eu},
M. Fouch{\'e}$^{1}$,
G. Labeyrie$^{1}$,
D. Vernet$^{2}$,\newline
\newauthor
F. Vakili$^{2}$,
and R. Kaiser$^{1}$
\\
$^{1}$Universit{\'e} C{\^o}te d'Azur, CNRS, INPHYNI, France\\
$^{2}$Universit{\'e} C{\^o}te d'Azur, OCA, CNRS, Lagrange, France}

\date{Accepted XXX. Received YYY; in original form ZZZ}

\pubyear{2018}

\begin{document}
\label{firstpage}
\pagerange{\pageref{firstpage}--\pageref{lastpage}}
\maketitle

\begin{abstract}
The present article reports on the first spatial intensity interferometry measurements on stars since the observations at Narrabri Observatory by Hanbury Brown \textit{et al.} in the 1970's. Taking advantage of the progresses in recent years on photon-counting detectors and fast electronics, we were able to measure the zero-time delay intensity correlation $g^{(2)}(\tau = 0, r)$ between the light collected by two $1${-m} optical telescopes separated by $15$~m. Using two marginally resolved stars ($\alpha$~Lyr and $\beta$~Ori) with R~magnitudes of $0.01$ and $0.13$ respectively, we demonstrate that $4$-hour correlation exposures provide reliable visibilities, whilst a significant loss of contrast is found on $\alpha$~Aur, in agreement with its binary-star nature.
\end{abstract}

\begin{keywords}
techniques: interferometric
\end{keywords}

%%%%%%%%%%%%%%%%%%%%%%%%%%%%%%%%%%%%%%%%%%%%%%%%%%

%%%%%%%%%%%%%%%%% BODY OF PAPER %%%%%%%%%%%%%%%%%%

%%%%%%%%%%%%%%%%%%%%%%%%%%%%%%%%%%%%%%%%%%%%%%%%%%%%%%%%%%%%%
\section{Introduction}

Surpassing the angular resolution of the present ground-based largest optical telescopes, \emph{i.e.} $8$ to $10$~m, in spite of the atmospheric turbulence, remains a challenge of modern observational  astrophysics. Future $30$-$40$~m extremely large telescopes, even equipped with laser-guided adaptive optics, will still be limited typically to $10$~mas resolution in the visible. Aperture synthesis techniques are nowadays currently used to observe mas structures on the surface or around stars in the visible~\citep{Mourard:2015,Garcia:2016,Gomes:2017}. These amplitude interferometers are very sensitive and have produced unique astrophysical results in the last decades. However, they require a real-time control of the optical path difference (OPD) between the telescopes below the optical wavelength, a daunting task if the baseline is to be further extended to the km range and with increasing difficulty at short wavelengths and for faint stars.

An alternative solution is ``intensity interferometry'', a technique based on the measurement of the intensity correlation function between light collected by several telescopes from a same source. This technique was successfully employed by Hanbury Brown and Twiss (HBT) in the 60's and 70's~\citep{HBT:1956,HBT:1974}, but latter abandoned due to its intrinsic lack of sensitivity in terms of limiting magnitude. However, intensity interferometry presents the important advantage to be easier to implement, since the acceptable fluctuations of the OPD are not determined by the optical wavelength but by the temporal resolution of the detection chain\,: a typical resolution of $100$~ps corresponds to a maximal OPD variation of $3$~cm. As a consequence, intensity interferometers are essentially insensitive to atmospheric turbulence for detectors with temporal resolutions above a few ps~\citep{HBT:1958,Tan:2016}. This fact, as well as the progress achieved in the last decades in the quantum efficiency of detectors and down-stream digital electronics, have sparked a renewed interest in intensity interferometry involving several research groups worldwide~\citep{Horch:2013,Dravins:2016,Zampieri:2016,Tan:2016,Pilyavsky:2017,Wang:2018,Matthews:2018,Rivet:2018}.

Recently, we reported the first measurement of the temporal intensity autocorrelation function from an unresolved star since the seminal work of HBT~\citep{Guerin:2017}. The present article presents the first re-implementation of a two-telescope intensity interferometer, accessing the spatial intensity correlation function from two $1${-m} optical telescopes separated by an East-West baseline of $15$~m. The first demonstrations were done on two marginally resolved stars, $\beta$~Ori (Rigel) and $\alpha$~Lyr (Vega), to check the reliability of our setup. We also observed $\alpha$~Aur (Capella), for which a loss of mean visibility is clearly measured, consistent with the binary nature of this object. In all cases the signal-to-noise ratio (SNR) of our measurements of the intensity correlation function is limited by the photon statistics.

%%%%%%%%%%%%%%%%%%%%%%%%%%%%%%%%%%%%%%%%%%%%%%%%%%%%%%%%%%%%%
\section{Setup}\label{setup}

\subsection{Principle}

To measure spatial intensity correlations, one needs to correlate intensities collected by two telescopes distant by $r$. The quantity of interest is the time correlation function given by
\begin{equation}
g^{(2)}(\tau, r) = \frac {\left\langle I(t, 0)I(t+\tau, r)\right\rangle}{\left\langle I(t, 0)\right\rangle \left\langle I(t, r)\right\rangle},
\end{equation}
where the brackets denote the average over times $t$.

In ref.~\citep{Guerin:2017}, we used a single telescope and thus measured the temporal autocorrelation function $g^{(2)}(\tau, r=0)$. The shape of this function allows to identify the nature of the light source. For instance, a coherent field such as emitted by a laser yields $g^{(2)}(\tau) = 1$ for any $\tau$. Thermal (chaotic) light with a coherence time $\tau_\mathrm{c}$ (inversely proportional to the spectral bandwidth $\Delta \nu$), such as detected from a star, yields a correlation function peaking at $2$ for $\tau = 0$ and decaying to $1$ for $\tau \gg \tau_\mathrm{c}$. The contrast of the correlation function (also called the ``bunching peak'') is given by $C = g^{(2)}(\tau = 0) - 1$. For thermal and spatially unresolved light, $C=1$. In situations where the coherence time can be resolved, the temporal correlation function can inform on the processes taking place in the light source, such as the temperature of a cold-atom sample~\citep{Nakayama:2010,Eloy:2018} or multiple scattering processes in a hot vapor~\citep{Dussaux:2016}.

When the source has a broad spectrum, however, the temporal resolution of the detection chain $\tau_\mathrm{det}$ is usually lower than $\tau_\mathrm{c}$ (i.e. $\tau_\mathrm{c}\ll\tau_\mathrm{det}$), in which case the correlation function is convoluted with the temporal response of the detection, yielding a bunching peak with a width set by $\tau_\mathrm{det}$. This leads to a reduced contrast $C \simeq \tau_\mathrm{c}/\tau_\mathrm{det}$, which has to be calibrated, either in the lab or on sky, with an unresolved (natural or
artificial) star.

%When two telescopes at different positions are used, one can reconstruct a visibility curve $V(r)$ and infer the source's angular diameter from the evolution of $g^{(2)}(\tau = 0, r)$ as the on-sky projected baseline length $r$ varies~\citep{HBT:1956}. For a uniform disk of angular diameter $\theta$ and assuming that the source is not resolved by a single telescope, one gets\,\citep{HBT:1958}:
%\begin{equation}
%V(r) = g^{(2)}(\tau = 0, r) - 1 = C \left[\frac{2 \mathrm{J}_1(x)}{x}\right]^2, \label{eq.g2r}
%\end{equation}
%with $x = \pi r\theta/\lambda_0$ and $\lambda_0$ the central wavelength.

\subsection{Instrumental setup}

\begin{figure}
    \includegraphics[width=\columnwidth]{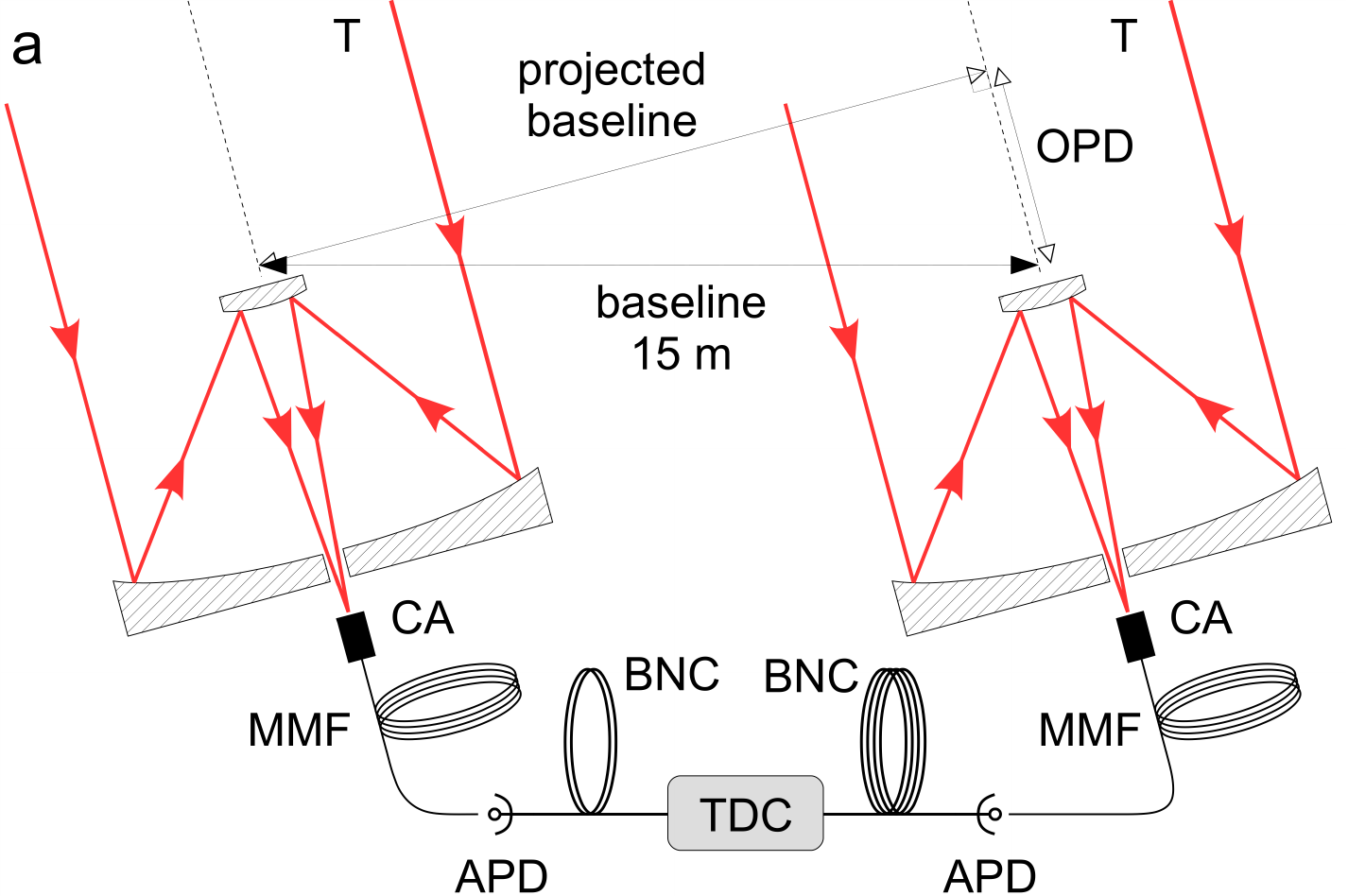}
    \caption{Experimental setup. The star light collected by two similar 1-m telescopes (T) separated by a 15-m baseline is coupled by the coupling assemblies (CA), into two multimode optical fibers (MMF) and detected by two avalanche photodiodes (APDs), connected via coaxial cables to a time-to-digital convertor (TDC). A CCD camera is also connected to each CA in order to track the desired star.}
    \label{fig.Setup}
\end{figure}

To achieve the measurements presented in this article, we used the two $1$-m telescopes of the C2PU facility at the plateau de Calern site of Observatoire de la C{\^o}te d'Azur as illustrated in Fig.~\ref{fig.Setup}. These two telescopes are separated by a fixed, nearly East-West baseline of $15$~m and their absolute positions are known through GPS differential measurements with $2\sigma$ uncertainties of $\pm1$~cm in the horizontal plane and $\pm2$~cm along the vertical direction. Coupling assemblies (CAs) collect, filter and inject the light from the telescopes into optical fibers, and allow monitoring the star's position on auxiliary cameras. The CA was presented in~\citep{Guerin:2017}. In summary, it is composed of two cascaded focal reducers to decrease the focal spot size at the position of the fiber tip, plus a polarizer and two frequency filters of $10$~nm and $\Delta \lambda=1$~nm bandwidths respectively. The role of the spectral filters is to increase the width of the intensity correlation function $\tau_c \sim \lambda_0^2/(2c\Delta\lambda) \simeq 1$~ps, where $c$ is the speed of light and $\lambda_0 = 780$~nm is the central wavelength of the filters. The light collected by each telescope is then injected in a $20$~m-long multimode optical fiber (MMF) connected to an avalanche photodiode (APD). The effective temporal resolution of the APDs is of the order of $500$~ps (FWHM).
%Since this time is much larger than $\tau_\mathrm{c}$, it determines the width of the bunching peak. As discussed before, increasing $\tau_\mathrm{c}$ increases the contrast of the bunching peak. On the other hand, if we are shot noise limited, one can show that the signal to noise ratio (SNR) is independent of $\tau_\mathrm{c}$ (and thus $\Delta \lambda$)\,\citep{HBT:1967a}. % Not very useful...
Finally, the outputs of the APDs are sent to a time-to-digital convertor (TDC) through $50\,\Omega$ coaxial cables. To avoid the cross-talk of the TDC around zero delay between channels, we introduce a large electronic delay of about $200$~ns by using different lengths for the two coaxial cables ($2$~m and $45$~m respectively).

The experiment described in this article was performed using the ``Time Tagger Ultra'' TDC from Swabian Instruments. Compared to the TDC used in \citep{Guerin:2017}, it presents several improvements. First, the true temporal correlation function is computed and recorded in real time, as well as the total photon flux. Second, the correlation artifact described in~\citep{Guerin:2017} is absent, eliminating the need to subtract the correlation background from a ``white'' source. The timing resolution of this TDC is $29$~ps (FWHM) and is not a limitation for this experiment since it is much better than the resolution of the APDs.

\begin{table*}
\caption{Main characteristics of the three stars used for this experiment.}
\label{tab.Stars}
\begin{tabular}{lllllll}
\hline
Name & $\alpha_{2000}$ & $\delta_{2000}$ & R mag & I mag & Spectral type & Comment \\
\hline
$\beta$~Ori & $05^h\,14^m\,32.27^s$ & $-08^\circ\,12'\,05.9''$ & +0.13 & +0.15 & B8Iae & Blue supergiant\\

$\alpha$~Lyr  & $18^h\,36^m\,56.34^s$ & $+38^\circ\,47'\,01.3''$ & +0.07 & +0.10 & A0Va &  \\

$\alpha$~Aur & $05^h\,16^m\,41.36^s$ & $+45^\circ\,59'\,52.8''$ & -0.52 & -0.96 & G8III + G0III & Spectroscopic binary \\
\hline
\end{tabular}
\end{table*}

\subsection{Visibility measurement} \label{subsec:V_meas}

The squared visibility $V^2$ is given by the height of the temporal correlation function extracted using a Gaussian fit. However, during the tests performed in the laboratory using an unresolved artificial star~\citep{Guerin:2017}, we found small day to day fluctuations of the bunching peak. This hints to an additional, fluctuating source of convolution for the correlation peak, inducing a fluctuation of the detection resolution $\tau_\mathrm{det}$ and thus of the contrast $C$. The origin of this fluctuation is at present not completely understood, but
%might be attributed to the distortions of the APD-generated pulses in the long coaxial cable. It is thus more reliable to measure the area of the bunching peak than its height. We note that
this problem can be overcome by measuring the area of the bunching peak instead of its height.
This area $A$, expressed in ps, does not depend anymore on $\tau_\mathrm{det}$ and gives access to the correlation time $\tau_\mathrm{c}$, with $A \simeq \tau_\mathrm{c}$,
with a multiplication factor which depends on the precise spectral shape of the filter.
Laboratory measurements of $g^{(2)}(\tau,0)$ on an artificial star gives $A_0 =1.37 \pm 0.05$~ps, which serves as zero baseline calibration for following on-sky observations.
%A measurement performed with our setup in the laboratory with an unresolved artificial star gives $A = 1.37 \pm 0.05$~ps. This value will thus serve as the reference corresponding to maximal visibility.

%We stress that, using the area of the bunching peak instead of its value at zero delay not only allows us to get rid of
%the time jitter at the detection, but it also provides a measurement of the coherence time of light. In our case, this %coherence time is determined by the width of the optical filter, but this method of data analysis could be useful when %looking for narrow emission lines of astrophysical sources.
% --> not very useful here, we'll keep this idea for a later experiment

%%%%%%%%%%%%%%%%%%%%%%%%%%%%%%%%%%%%%%%%%%%%%%%%%%%%%%%%%%%%%
\subsection{Data acquisition and analysis}
\label{TDC}

As can be seen from Fig.~\ref{fig.Setup}, the position of the star in the sky determines the on-sky projected baseline (which varies between $9.6$ and $15$~m in our observations), and the optical path difference (OPD) between the two telescopes. This later quantity is important since it introduces an optical time delay $t_\mathrm{d} = \mathrm{OPD}/c$. During the full observation period this delay can vary up to $38$~ns, corresponding to an OPD variation of $11.5$~m. %In addition, there is a fixed electronic delay of about $200$~ns introduced by the length difference in the coaxial cables connecting the APDs to the TDC. As mentioned before, the purpose of this delay is to shift the correlation peak away from the zero delay region where the TDC introduces spurious correlations.

During the observation nights, we measured the correlation signal in real time and recorded it every $10$~seconds. The knowledge of the absolute recording time and the star position allows us to compute both the projected baseline and the OPD for each recording period. The averaging over a given number of recordings is then performed by first time-shifting every recording by the corresponding optical delay $t_\mathrm{d}$, and then adding up all the shifted recordings. The electronic delay is also subtracted.

%%%%%%%%%%%%%%%%%%%%%%%%%%%%%%%%%%%%%%%%%%%%%%%%%%%%%%%%%%%%%
\section{Results and Discussion}
\label{results_discussion}

\subsection{Observation conditions}
\label{stars}

During the four nights from 10/10/2017 to 10/14/2017, we observed three bright stars: $\beta$~Ori, $\alpha$~Lyr, and $\alpha$~Aur. Their main parameters are summarized in Table~\ref{tab.Stars}. The observational conditions are detailed in Table~\ref{tab.ObsRuns}.

The stars $\beta$~Ori and $\alpha$~Lyr have angular diameters of respectively $\theta = (2.526 \pm 0.006)$~mas and $\theta = (3.198 \pm 0.016)$~mas (uniform disk diameters,~\cite{Baines:2018}), which correspond to baselines of $77.7$~m and $61.4$~m to be fully resolved at $\lambda_0=780$~nm (first minimum of the spatial correlation function given by Eq.~(\ref{eq.g2r})). Since our maximal baseline is $15$~m, the expected loss of squared visibility is at most $13\%$ for $\beta$~Ori and $20\%$ for $\alpha$~Lyr. Thus, these stars were used as ``calibration'' sources to check the consistency of our observations and the reliability of our detection method. On the contrary, $\alpha$~Aur is a binary system whose individual components have roughly twice the apparent diameter of $\alpha$~Lyr or $\beta$~Ori. In addition, the binary nature of the star system leads to oscillations of the visibility during the observation time, as will be further discussed in section~\ref{results_Capella}.

\begin{table}
\caption{Main circumstances for the observing runs performed on the three stars. Begin and end dates are in UTC (ISO~8601 compact format). $a$ is the air mass range. The seeing information is provided by the GDIMM instrument
\citep{Ziad:2012,Aristidi:2014} of the CATS station (Calern Atmospheric Turbulence Station) \citep{Chabe:2016}. The numbers are median values over the whole nights.}
\label{tab.ObsRuns}
\renewcommand{\tabcolsep}{2pt}
\begin{tabular}{lcccc}
\hline
Star& Begin & End & $a$ & Seeing\\ %, date
\hline
$\beta$~Ori\,  & $20171011T0033Z$\, & $20171011T0455Z$\, &
$1.62\rightarrow2.61$\, & $0.91^{\prime\prime}$\\    %10 - 11 Octobre 4h22 (4.4h)
$\alpha$~Lyr\, & $20171011T1803Z$\, & $20171011T2207Z$\, & $1.02\rightarrow1.88$\, & $1.86^{\prime\prime}$\\    %11 - 12 Octobre 4h04 (4.0h)
$\alpha$~Lyr\, & $20171012T1814Z$\, & $20171012T2203Z$\, & $1.03\rightarrow1.96$\, & $1.10^{\prime\prime}$\\    %12 - 13 Octobre 3h49 (3.8h)
$\alpha$~Aur\, & $20171012T2247Z$\, & $20171013T0514Z$\, & $1.00\rightarrow1.61$\, & $1.10^{\prime\prime}$\\    %12 - 13 Octobre 6h27 (6.5h)
$\alpha$~Lyr\, & $20171013T1836Z$\, & $20171013T2159Z$\, & $1.05\rightarrow2.00$\, &         n.a.\\    %13 - 14 Octobre 3h23 (3.4h
$\alpha$~Aur\, & $20171013T2235Z$\, & $20171014T0507Z$\, & $1.00\rightarrow1.58$\, &         n.a.\\    %13 - 14 Octobre 6h32 (6.5h)
\hline
\end{tabular}
\end{table}

\subsection{Results on marginally resolved stars}\label{results_unresolved}

\begin{figure*}
    \begin{subfigure}{0.66\columnwidth}
        \centering \includegraphics[width=\columnwidth]{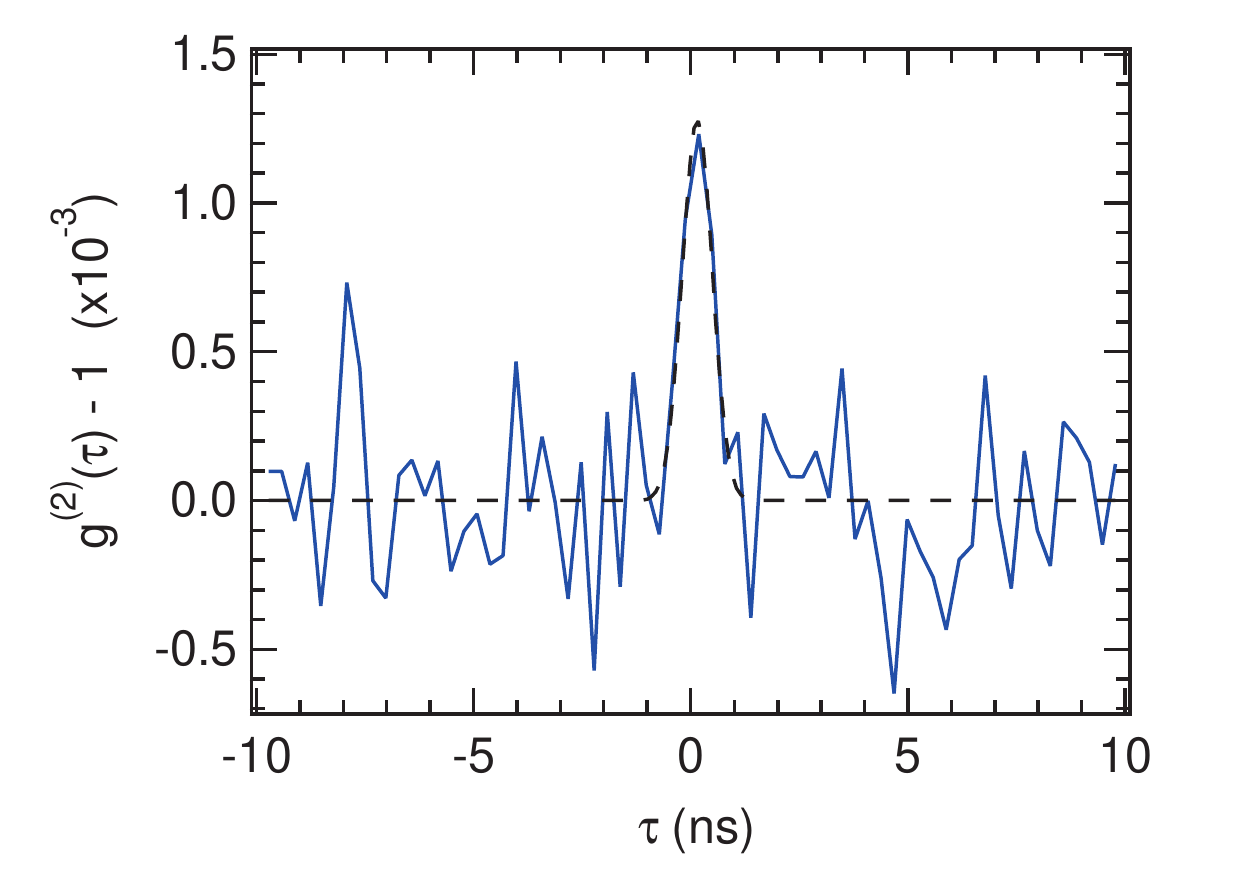}
        \caption{$\beta$~Ori.}
        \label{fig.g2a}
    \end{subfigure}
    \begin{subfigure}{0.66\columnwidth}
        \centering \includegraphics[width=\columnwidth]{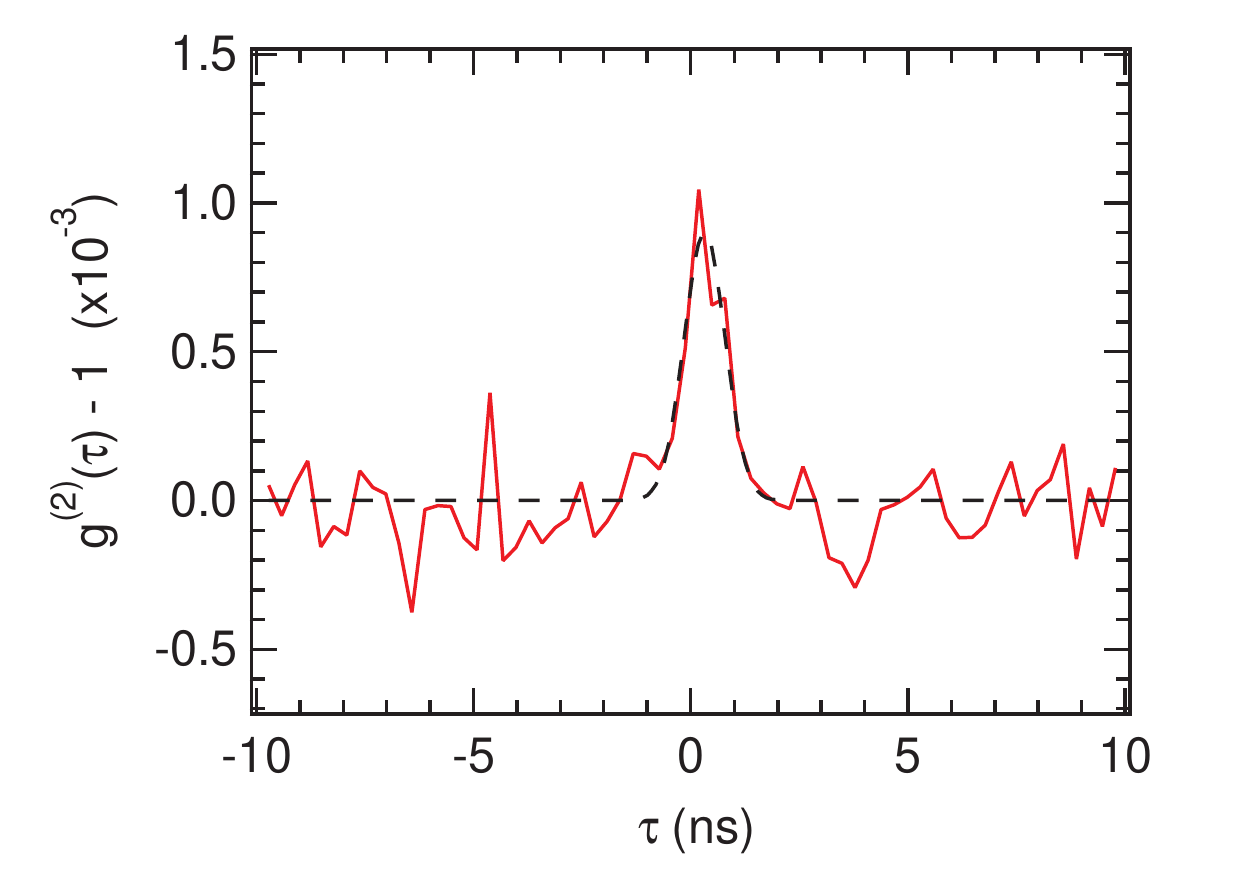}
        \caption{$\alpha$~Lyr.}
        \label{fig.g2b}
    \end{subfigure}
    \begin{subfigure}{0.66\columnwidth}
        \centering \includegraphics[width=\columnwidth]{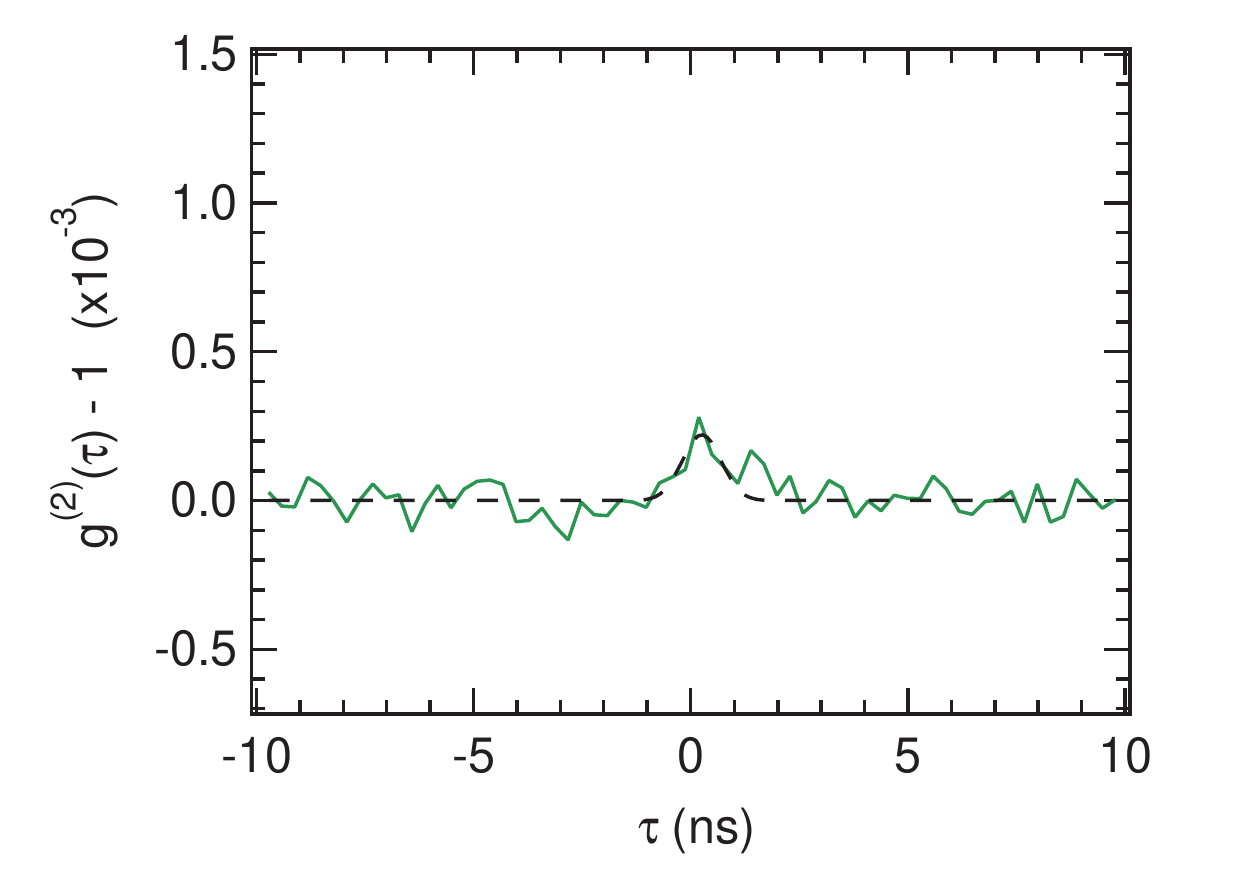}
        \caption{$\alpha$~Aur.}
        \label{fig.g2c}
    \end{subfigure}
    \caption{Temporal correlation functions for two marginally resolved stars and one binary star (solid line\,: data, dashed line: Gaussian fit). a) $\beta$~Ori. The integration time is $4.3$~h. The average count rate per detector is $1.8$~Mcounts/s. The projected baseline varied between $10.7$ and $15$~m.  b) $\alpha$~Lyr. The total integration time is $11.1$~h over three nights. The average count rate per detector is $2.3$~Mcounts/s. The projected baseline varied between $9.6$ and $14.3$~m.  c) $\alpha$~Aur. The total integration time is $12.5$~h over two nights. The average count rate per detector is $4.9$~Mcounts/s. The projected baseline varied between $11.5$ and $15$~m.}
\label{fig.g2_r}
\end{figure*}

We first present in Figs.~\ref{fig.g2a} and \ref{fig.g2b} the correlation curves obtained for $\beta$~Ori and $\alpha$~Lyr. As these stars are not resolved and the projected baseline variation is rather small, we show the data averaged over the full observation duration ($4.3$~h for $\beta$~Ori and $11.1$~h over three successive nights for $\alpha$~Lyr). The bunching peak is clearly seen for both stars.

\begin{figure}
    \includegraphics[width=\columnwidth]{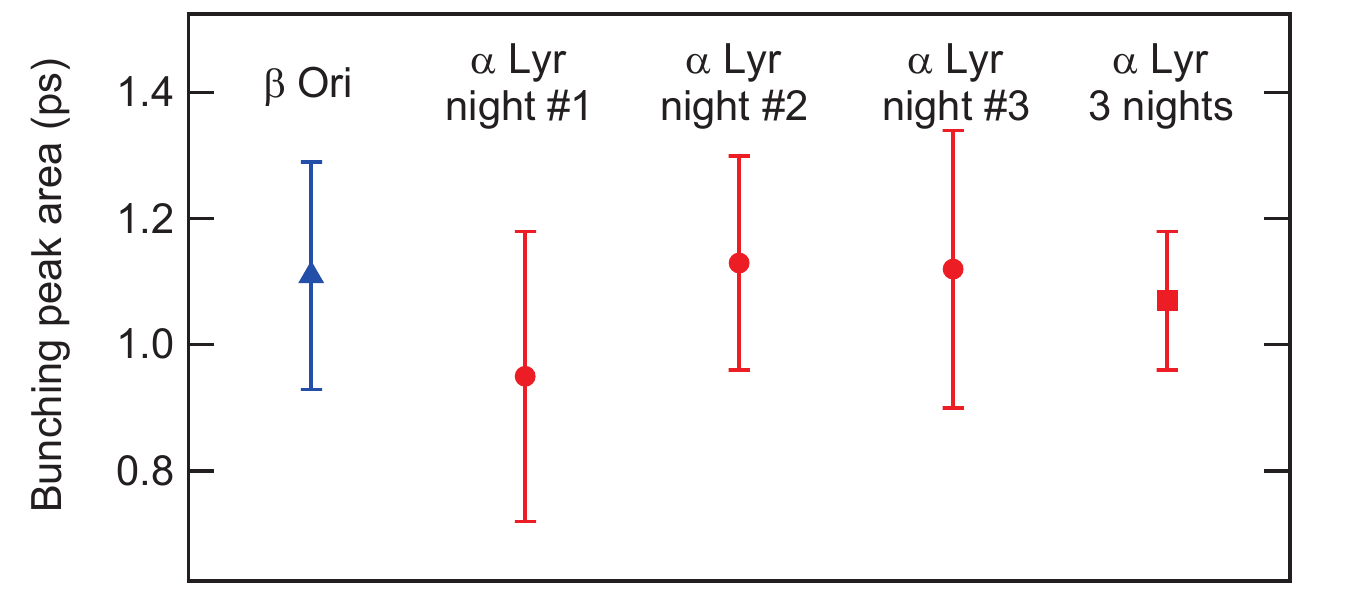}
    \caption{Area of the photon bunching peak in ps for $\beta$~Ori and $\alpha$~Lyr for the different nights of observation. For each night the observation time is around $4$~h. The error bars correspond to the $1\sigma$ uncertainty. All the values are consistent with each other within the error bars.
%\textcolor{red}{\textbf{Cette facon de presenter les choses n\'est pas courante pour les astrophysiciens. Il est preferable de ramener cela a des visibilites, donc de 1 a 0 en ordonnee et en mas ou en (micro-)radian en abscisse}}.
}
\label{fig.Visibility_Vega_Rigel_nights}
\end{figure}

We observe that the bunching peak from $\beta$~Ori has a higher contrast than that from $\alpha$~Lyr, whilst its width is smaller. By comparing the three curves obtained for each night of observation on $\alpha$~Lyr, we also found some fluctuations of the height and width of the correlation peaks. As discussed in section\,\ref{subsec:V_meas}, this phenomenon was also observed in the laboratory and we concluded that the measurement of the bunching peak area is more reliable than its amplitude at zero delay. As illustrated in Fig.~\ref{fig.Visibility_Vega_Rigel_nights}. The measured area for the two stars and for different nights of observation are well consistent with each other.

\begin{figure}
    \includegraphics[width=\columnwidth]{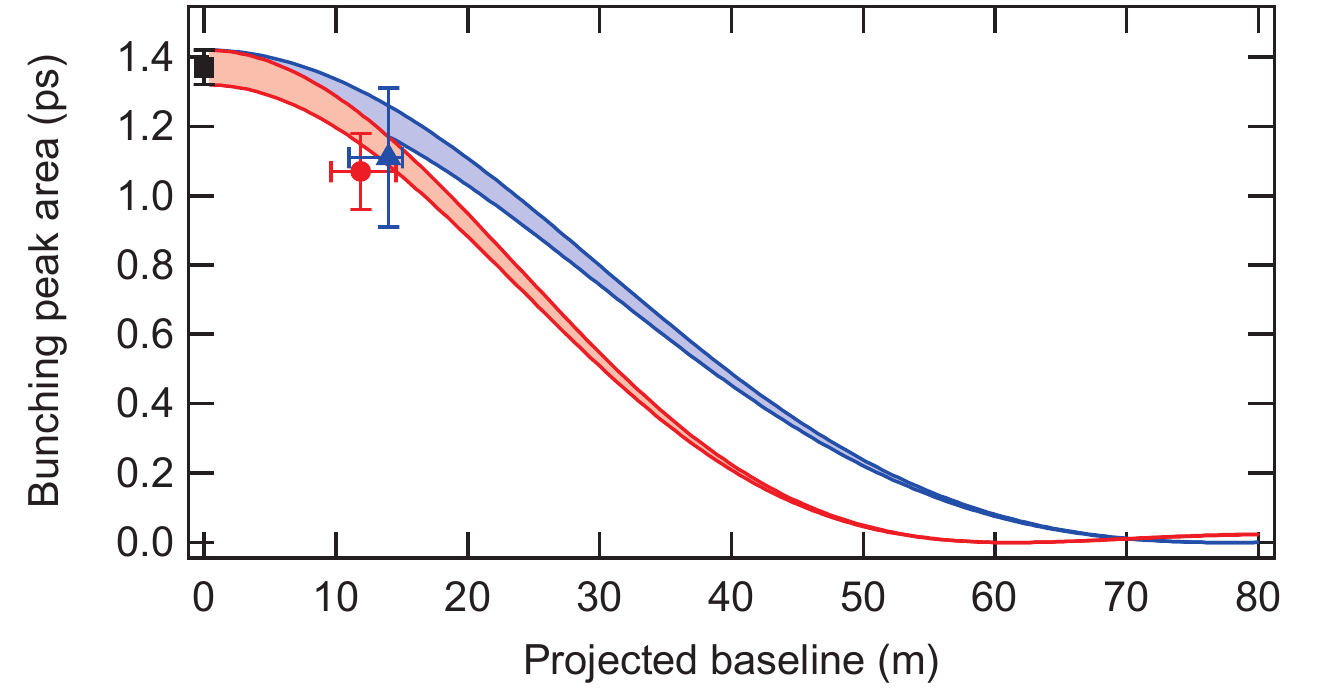}
    \caption{Area of the photon bunching peak for $\alpha$~Lyr (red circle) and $\beta$~Ori (blue triangle) with their respective error bars ($A=1.07\pm0.11$~ps and $A=1.11\pm0.20$~ps respectively). The black square at zero baseline corresponds to a calibration of $g^{(2)}(\tau,0)$ performed in the laboratory using an artificial star ($A_0=1.37\pm0.05$~ps). The red and blue pairs of curves and shaded areas in between indicate the expected visibility curves for $\alpha$~Lyr and $\beta$~Ori respectively.}
    \label{fig.Visibility_Vega_Rigel}
\end{figure}

We can now compare in Fig.~\ref{fig.Visibility_Vega_Rigel} our data to the expected visibility curves of $\beta$~Ori and $\alpha$~Lyr calculated using Eq.~(\ref{eq.g2r}) and their known angular diameters \citep{Baines:2018}. On this plot, the black square at zero baseline corresponds to the calibration of $g^{(2)}(\tau,0)$ performed in the laboratory, $A_0 =1.37 \pm 0.05$~ps (section~\ref{subsec:V_meas}). This value is used to normalize the computed squared visibility. The triangle and circle correspond to the fully averaged data for $\beta$~Ori and $\alpha$~Lyr respectively. The horizontal error bars correspond to the variation of the projected baseline during the total observation time. The vertical error bars reflect the $1\sigma$ uncertainty on the area extracted from the fits.
The solid lines surrounding the shaded areas correspond to the theoretical uniform disk visibility curves for the two stars, given by
\begin{equation}
V^2(r) = \left[\frac{2 \mathrm{J}_1(\pi r\theta/\lambda_0)}{\pi r\theta/\lambda_0}\right]^2. \label{eq.g2r}
\end{equation}
For each pair, the upper and lower lines are normalized with the upper and lower bounds of the calibrated $A_0$.
We see a good agreement between the measurements and the expected curves, without any free parameter, showing that we can either use the laboratory measurement or the measurements on unresolved stars for the normalization.

The ratio between the measured visibility and its 1$\sigma$ uncertainty defines the SNR of the measurement. It is $10$ for $\alpha$~Lyr and $6$ for $\beta$~Ori, due to a shorter integration time. In both cases it is well consistent by the limit $SNR = C \sqrt{N_\mathrm{c}}$ given by the contrast $C$ and the finite number of coincidences $N_\mathrm{c}$ accumulated during the integration time $T$ in a time window $\tau_\mathrm{det}$, $N_\mathrm{c}=F^2T\tau_\mathrm{det}$, with $F$ the detected photon flux per detector. This shows that no spurious correlation degrades the performance of our instrument. This is an important advantage of working in the photon-counting regime.

\subsection{Results on $\alpha$~Aur}\label{results_Capella}

As to the observations on $\alpha$~Aur, the data was recorded during two nights (total duration of $12.5$~h), with a high photon count rate of $4.9$~Mcounts/s per detector. Fig.~\ref{fig.g2c} shows the correlation peak corresponding to an averaging over the full dataset. As can be seen, the contrast of the bunching peak is strongly reduced but still measurable. The area of the bunching peak is now $(0.25 \pm 0.05)$~ps, about $4$ times lower than for $\alpha$~Lyr and $\beta$~Ori. This observation is an unambiguous signature that $\alpha$~Aur as a whole is resolved.

\begin{figure}
    \includegraphics[width=\columnwidth]{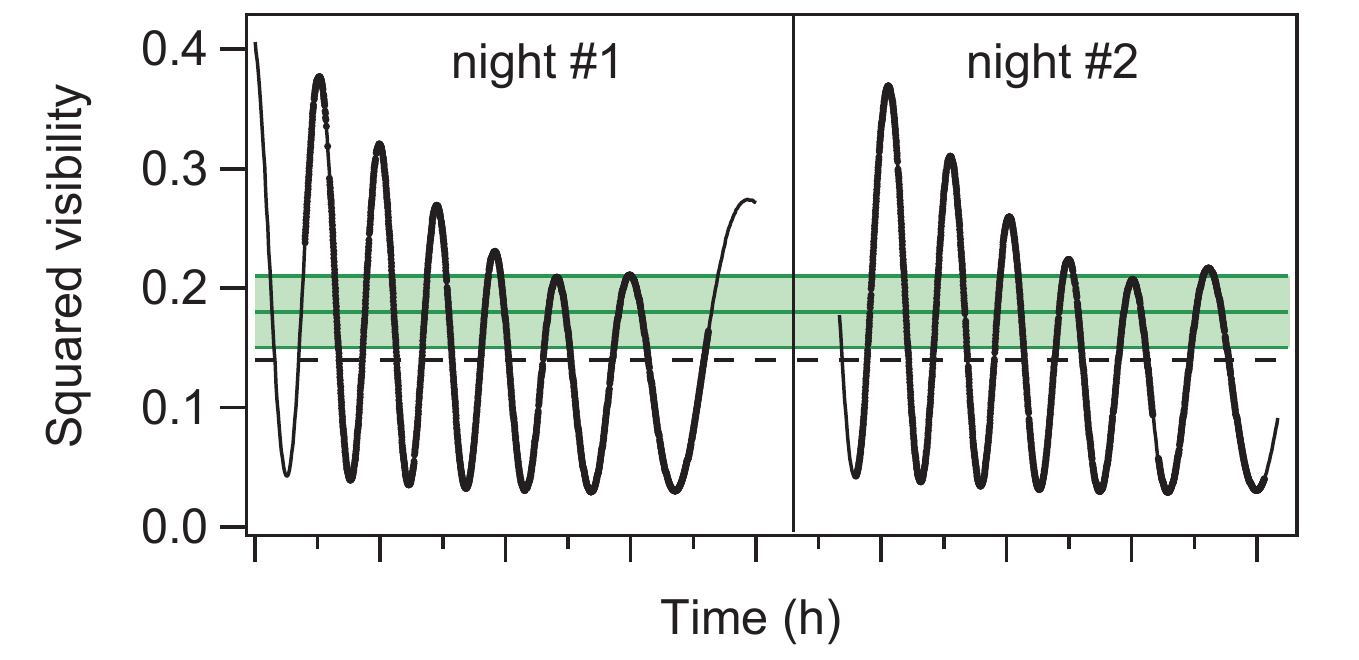}
    \caption{Expected normalized squared visibility of $\alpha$~Aur during the two nights of observations. The continuous curves are computed by Eq.~(\ref{Eq.V_binary}) convoluted by the pupil shape of each telescope, and black dots correspond to instants where data have been recorded. The horizontal dashed line show the averaged computed visibility for the observation. The measured visibility is represented in green as a 68\% confidence interval.}
    \label{fig.Visibility_Capella}
\end{figure}

However, the detailed analysis is complicated by the fact that $\alpha$~Aur is a binary source. The visibility now depends not only on the projected baseline but also on the angular distance between the binary components and the angle they make with the on-sky projection of the baseline\,\citep{HBT:1967b}\,:
\begin{multline}
V^2(r) = \frac{1}{(I_a+I_b)^2} \times
\biggl[I_a^2 V_a^2(r) + I_b^2V_b^2(r)\\
+ 2I_aI_bV_a(r)V_b(r) \cos\Bigl(\frac{2\pi r\overline{\theta}\cos\psi}{\lambda_0} \Bigr)\biggr],
\label{Eq.V_binary}
\end{multline}
with $I_a$ and $I_b$ the intensities of the A and B components at $780$~nm, $V_a(r)$ and $V_b(r)$ their respective visibility curves, $r$ the on-sky projection of the baseline, $\overline{\theta} \simeq 55$~mas the angular separation between the two components and $\psi$ the angle between the on-sky projected baseline and the vector separation of the two components.
The visibility is expected to oscillate as a function of time, mainly because of the daily rotation of the on-sky projected baseline \emph{w.r.t.} the $A-B$ direction of the binary star. Since $\alpha$~Aur's individual components have nearly identical luminosities [$I_1=(84.5\pm8.6)\times L\odot$ and $I_2=(76.3\pm6.3)\times L\odot$ at $780\pm0.5$~nm\,; estimation through a black body model with physical parameters from \cite{Torres:2015}], the contrast of the visibility oscillations is large, as shown in Fig.~\ref{fig.Visibility_Capella} for the two observation nights. Note that Eq.~(\ref{Eq.V_binary}) is only valid when each telescope is too small to resolve the binary star, which is presently not completely the case. In particular, this induces a reduction of the correlation at zero baseline~\citep{HBT:1958II}. This effect is taken into account for the calculations considering the size of the telescope as well as its central obstruction (approximated to $30$~cm in diameter).

Although the SNR is also limited by the coincidence statistics, it
is too low to extract in a reliable way the evolution of the visibility as a function of time as it would need to use integration times of a few minutes. We can only compare the expected and measured visibilities averaged over the whole observing period. We get $\langle V^2_\mathrm{mes} \rangle=0.18\pm0.03$ for the measured averaged normalized squared visibility, in fair agreement with the computed one $\langle V^2_\mathrm{th} \rangle=0.14$.

This puts in evidence a limitation of intensity interferometry in cases with the visibility of the source varies in time. In their first study of a binary source at Narrabri ($\gamma^2$\,Vel), Hanbury Brown \textit{et al.} also only report an average visibility as a function of the baseline without resolving the faster oscillations~\citep{HBT:1970}. In a subsequent study of $\alpha$\,Vir they exploited the fact that the binary orbital period is a quasi-integer number of days (4.014, already known) to integrate data over several nights and obtain a visibility curve for each of the four accessible phases of the orbital period~\citep{HBT:1971}.

%%%%%%%%%%%%%%%%%%%%%%%%%%%%%%%%%%%%%%%%%%%%%%%%%%%%%%%%%%%%%
\section{Conclusion}

We have reported in this article the first measurements of intensity correlations from two $1${-m} telescopes in the visible domain, since the Narrabri Observatory measurements achieved by Hanbury Brown \textit{et al.} 50 years ago. We show that in spite of the modest collecting surface of our telescopes and of our limited observation time (typically a few hours per target star), we measured the contrast of the bunching peak, alternatively the squared visibility, with a SNR ratio on the order of $10$ on two early type stars\,: $\alpha$~Lyr and $\beta$~Ori of spectral types A and B respectively. With the bright binary G type star $\alpha$~Aur, we have been able to measure a squared visibility as low as $14\%$, corresponding to the partial resolution of the individual components of the binary combined with min-max modulation of the visibility.

The present demonstration of intensity interferometry with rather modest telescopes
shows the power of modern photon-counting technologies to perform shot-noise limited measurements and
also confirms that intensity interferometry can be extended nowadays to red and IR wavelengths, probably to the H spectral band.

Building upon these results we plan to use such two-telescope correlation measurements to perform polarization-resolved intensity interferometry on bright stars. Such measurements have proven extremely difficult with amplitude interferometry, mainly due to the complexity of the optical path involving time-variable oblique reflections introducing instrumental polarizations. Interferometry on spectral emission line, especially at short wavelengths, is also a scientific niche where intensity interferometry should play a major role.

Finally, today's time-transfer technologies enable the synchronization of separate telescopes as distant as several kilometers across. As demonstrated in this work, the focal instrument of individual telescopes can be directly replicated on as many telescopes as available and, thanks to the improving time resolution of detectors and backbone digital correlators, there is no limitation to extend baselines to several kilometers if not tens of kilometers. Alas, the longer the baselines, the lower the visibilities (or correlation contrast). Also, rapidly rotating multiple-star systems lead to varying visibility when the binary separation is much larger than the interferometric resolution, which limits the possible integration time. Therefore the question of improving the SNR becomes central. As suggested originally by Hanbury Brown himself \citep{HBT:1968}, one can use many simultaneous spectral channels with a net gain of square root of their number. Recording several hundreds of spectral channels and the two polarization channels could gain 3 to 4 magnitudes. Then much larger telescopes in the 4--10~m class will bring 3--5 more magnitudes to cover a broad spectrum of stars in the H-R diagram or for their intrinsic physical nature.

%%%%%%%%%%%%%%%%%%%%%%%%%%%%%%%%%%%%%%%%%%%%%%%%%%%%%%%%%%%%%
\section*{Acknowledgements}

We thank A.~Eloy for participating to observations, A.~Dussaux for the design of the coupling assemblies, A.~Siciak for final test experiments in the lab, and O.~Lai for fruitful discussions. We acknowledge the CATS team (Calern Atmospheric Turbulence Station, https://cats.oca.eu/) for providing turbulence real-time measurements and Swabian Instruments for lending the TDC. This work is supported by the UCA-JEDI project ANR-15-IDEX-01 and the OPTIMAL platform.


%\bibliographystyle{mnras}
%\bibliography{HBT_paper_biblio}

%%%%%%%%%%%%%%%%%%%%%%%%%%%%%%%%%%%%%%%%%%%%%%%%%%

%%%%%%%%%%%%%%%%% APPENDICES %%%%%%%%%%%%%%%%%%%%%

%\appendix
%
%%If you want to present additional material which would interrupt the flow of the main paper,
%%it can be placed in an Appendix which appears after the list of references.
%

% Don't change these lines
\bsp    % typesetting comment
\label{lastpage}
\end{document}